\documentclass[twocolumn]{autart}    

\usepackage{graphicx}          

\usepackage{amsmath}
\usepackage{amssymb}
\usepackage{amsfonts}
\usepackage{mathrsfs}
\usepackage{xcolor}
\usepackage[ruled,norelsize]{algorithm2e}
\usepackage{graphicx}
\usepackage{textcomp}
\usepackage{mathptmx}
\usepackage{stmaryrd}
\usepackage{textcomp}
\usepackage{multirow}
\usepackage{booktabs}

\begin{document}

\begin{frontmatter}

\title{An iterative scheme for finite horizon model reduction of continuous-time linear time-varying systems} 


\author[Paestum]{Kasturi Das}\ead{kasturidas@iitg.ac.in},    
\author[Paestum]{Srinivasan Krishnaswamy}\ead{srinikris@iitg.ac.in},               
\author[Paestum]{Somanath Majhi}\ead{smajhi@iitg.ac.in}  

\address[Paestum]{Department of Electronics and Electrical Engineering, IIT Guwahati, Assam, India}  


\begin{abstract}                          %
This paper obtains the functional derivatives of a finite horizon error norm between a full-order and a reduced-order continuous-time linear time-varying (LTV) system. 
Based on the functional derivatives, first-order necessary conditions for optimality of the error norm are derived, and a projection-based iterative scheme for model order reduction of continuous-time LTV system is proposed. The iterative scheme is initialized with reduced-order models (ROMs) obtained by the finite horizon balanced truncation (FH BT) algorithm for LTV systems. Finally, the performance of the proposed scheme is tested with the help of two numerical examples.     
\end{abstract}
\end{frontmatter}

\section{Introduction}


Linear time-varying (LTV) models are required to model systems whose behaviours change with time. A typical example of such a system is a missile whose weight changes along its trajectory due to fuel expenditure \cite{evers1992ltv}. Other examples of such systems include robotic manipulators with deployable joints and flexible links, deployable space structures, bridges with crossing vehicles, etc.  

Complex time-varying systems require large LTV models to capture system dynamics accurately. However, simulating, analyzing and designing controllers for such large models is computationally intensive. The time-varying nature of the model parameters adds to the complexity. 
Further, for capturing transient responses, approximating unstable models, etc, one may want to approximate the large model for a limited time interval. There is, therefore, a need for finite horizon model order reduction algorithms of LTV models.


The finite horizon model order reduction problem for linear time-invariant (LTI) systems is well-studied. 
Algorithms such as the singular value decomposition (SVD)-based Time-Limited Balanced Truncation (TL-BT)  \cite{gawronski1990model} and the time-limited $H_2$ optimal methods proposed in \cite{das2023near},  \cite{goyal2019time}, and \cite{sinani2019h2} address this problem. The methods proposed in \cite{das2023near} and \cite{goyal2019time} are based on interpolation and Lyapunov-based time-limited $H_2$ optimality conditions, respectively. We refer to the latter method as the Time-Limited Two-Sided Iterative Algorithm (TL-TSIA) in this paper. \cite{das2022h} proposes a gradient-based time-limited $H_2$ optimal model order reduction (MOR) method. It is based on a closed-form expression of the gradients of the time-limited $H_2$ error norm.

The generalization of the TL-BT algorithm for LTV models is well-studied. This algorithm has been generalized for continuous-time LTV systems in \cite{sandberg2006case}, \cite{sandberg2004balanced}, \cite{verriest1983generalized} etc. and for discrete-time LTV systems in \cite{lall2003error}, \cite{sandberg2004balanced}. In this paper, the continuous-time generalization is referred to as FH BT. The error bounds for FH BT are obtained in \cite{sandberg2004balanced}. In addition, numerical issues related to applying FH BT to LTV systems are addressed in \cite{chahlaoui2005model,lang2016balanced}. 
In contrast to the generalizations of the TL-BT algorithm, the literature dealing with the generalization of the finite horizon $H_2$ optimal model order reduction problem for LTV systems is sparse. This problem has been addressed in \cite{melchior2012finite} and \cite{melchior2014model}, but exclusively for discrete-time LTV systems. 

 
To the best of the authors' knowledge, there are no finite horizon $H_2$ optimal model reduction algorithms for continuous-time LTV systems. The paper aims to address this research gap. The primary objective of this paper is to extend the TL-TSIA algorithm for continuous-time LTV systems.
The contributions of the paper can be summarized as follows. Firstly, we introduce a finite horizon $H_2$ error norm to measure the accuracy of reduced-order approximations of LTV systems. 
Then, we derive the necessary first-order conditions for the optimality of the error norm. 
Based on the optimality conditions, we propose a projection-based iterative scheme for finite horizon model order reduction of LTV systems. Finally, the effectiveness of the proposed iterative scheme is validated with the help of a couple of numerical examples.
The rest of the paper is arranged as follows. Section \ref{Section 2} presents some preliminary results related to 
time-limited $H_2$ optimal model reduction of continuous-time LTI systems. Section \ref{Section3} introduces an error norm for continuous-time LTV systems and proposes an iterative scheme for finite horizon $H_2$ optimal model order reduction of such systems. The performance of the proposed scheme is tested with the help of a numerical example in Section \ref{Section4}. The paper is concluded in Section \ref{Section5}. 

\noindent \textit{\underline{Notations}:
Let $\mathbb{R}$ be the set of real numbers and $[t_0,t_f]$ be a closed interval in $\mathbb{R}$. For a vector $v \in \mathbb{R}^n$, $\left\Vert v \right\Vert_2$ denotes the Euclidian norm. For a matrix $M \in \mathbb{R}^{n \times m}$, $M^T$ denotes the transpose, $\mathrm{Ran}(M)$ denotes the columnspace of $M$, $\left\Vert M \right\Vert_2$ denotes the induced 2-norm, $\left\Vert M \right\Vert_F$ denotes the Frobenius norm and $\left\Vert M \right\Vert_F^2 := \mathrm{Tr}(M^T M)$. Consider the continuous function $f:[t_0,t_f] \to \mathbb{R}^n$. The finite $L_2$ norm of $f$ is denoted by $\left\Vert f \right\Vert_{L_2^n[t_0,t_f]}:=\left( \int_{t_0}^{t_f} \left\Vert f(t) \right\Vert_2^2 dt \right)^{\frac{1}{2}}$. }

\section{Preliminaries}\label{Section 2}
This section briefly discusses the time-limited $H_2$ optimal MOR problem for LTI systems. Further, the Time-Limited Two-Sided Iterative Algorithm (TL-TSIA) is presented. 
\subsection{Time-limited $H_2$ optimal model order reduction (MOR)}\label{Sec2Subsec2_1}
The time-limited $H_2$ optimal MOR problem involves obtaining reduced-order LTI approximations of high-order LTI models, which are optimal with respect to a time-limited $H_2$ error norm. Consider, for example, an LTI system with the following state-space representation:
\begin{align}
    \dot{x}(t) &= \hat{A}x(t)+\hat{B}u(t), \nonumber \\
    y(t) &= \hat{C}x(t),  \label{LTIsystem1}
\end{align}
where $\hat{A} \in \mathbb{R}^{n \times n}$, $\hat{B} \in \mathbb{R}^{n \times m}$ and $\hat{C} \in \mathbb{R}^{p \times n}$. Next, we assume the following reduced-order LTI approximation of \eqref{LTIsystem1}. 
\begin{align} 
\dot{x}_r(t) &= \hat{A}_r {x_r}(t)+\hat{B}_r u(t),  \nonumber \\
y_r(t) &= \hat{C}_r x_r(t),  \label{red_LTIsystem1}
\end{align}
where $\hat{A}_r \in \mathbb{R}^{r \times r}$, $\hat{B}_r \in \mathbb{R}^{r \times m}$  and $\hat{C}_r \in \mathbb{R}^{p \times r}$ with $r \ll n$. The time-limited $H_2$ error norm between \eqref{LTIsystem1} and \eqref{red_LTIsystem1}  is defined as
$\left(\int_{0}^{\tau}\left\Vert \hat{C}e^{\hat{A}t}\hat{B}-\hat{C}_r e^{\hat{A}_r t} \hat{B}_r \right\Vert_F^2 dt\right)^{\frac{1}{2}}$ where $\tau < \infty$. The minimization of this error norm ensures that $y_r(t)$ is a good approximation of $y(t)$ over the time-interval $[0,\tau]$.
The Lyapunov-based conditions for optimality of the time-limited $H_2$ error norm are stated below. 
\begin{prop}\cite{das2022h}\label{LyapbasedOptim}
Let the reduced-order system (\ref{red_LTIsystem1}) be the best $r^{th}$ order approximation of the full-order system (\ref{LTIsystem1}) with respect to the time-limited $H_2$ error norm. Then,
\begin{align*}
   &Q_{r,\tau}\hat{P}_r-Y^{T}_\tau \hat{X}-\tau (L(\hat{A}_r \tau,S_{\tau})^T) = 0,\\
   &Q_{r,\tau}
   \hat{B}_r-Y^{T}_\tau \hat{B} = 0, \text{ and } \hat{C}_r P_{r,\tau}-\hat{C} X_\tau= 0.
\end{align*}
where the matrices $P_{r,\tau}$, $Q_{r,\tau}$, $X_{\tau}$, $Y_{\tau}$, $\hat{P}_r$ and $\hat{X}$ are solutions of the following matrix equations
\begin{align*}
    \hat{A}_r P_{r,\tau}+P_{r,\tau}{\hat{A}_r}^{T}+\hat{B}_r{\hat{B}_r}^{T}-e^{{\hat{A}_r}\tau}{\hat{B}_r}{\hat{B}_r}^{T}e^{{\hat{A}_r}^{T}\tau} = 0, \\
    {\hat{A}_r}^{T}Q_{r,\tau}+Q_{r,\tau}\hat{A}_r+{\hat{C}_r}^{T}{\hat{C}_r}-e^{\hat{A}_r^{T}\tau}{\hat{C}_r}^{T}{\hat{C}_r}e^{{\hat{A}_r}\tau} = 0,  \\
    \hat{A}X_\tau+X_\tau {\hat{A}_r}^{T}+\hat{B}{\hat{B}_r}^{T}-e^{\hat{A}\tau}\hat{B}{\hat{B}_r}^{T}e^{{\hat{A}_r}^{T}\tau} = 0 ,\\
    \hat{A}^{T}Y_\tau+Y_\tau \hat{A}_r+\hat{C}^{T}\hat{C}_r- e^{\hat{A}^{T}\tau}\hat{C}^{T}\hat{C}_r e^{\hat{A}_r\tau} = 0 ,
    \\
    \hat{P}_r \hat{A}_r^T + \hat{A}_r \hat{P}_r + \hat{B}_r \hat{B}_r^T = 0, \text{ and } \hat{X}^T \hat{A}^T + \hat{A}_r \hat{X}^T + \hat{B}_r \hat{B}^T = 0.
\end{align*}
Also, $L(\hat{A}_r \tau, S_{\tau}) = \int_{0}^{1} e^{\hat{A}_r \tau(1-s)}S_{\tau}e^{\hat{A}_r \tau s}ds$ is the Frechet derivative of the matrix exponential of $\hat{A}_r \tau$ along the direction $S_\tau=\left(\hat{X}^Te^{\hat{A}^T\tau}\hat{C}^T \hat{C}_r+\hat{P}_r e^{\hat{A}_r^T \tau}\hat{C}_r^T\hat{C}_r \right)$. 
\end{prop}
\subsection{Time-Limited Two-Sided Iterative Algorithm (TL-TSIA)}\label{Sec2Subsec2_2}
The TL-TSIA algorithm is based on the optimality conditions presented in Proposition \ref{LyapbasedOptim} \cite{goyal2019time}. It attempts to obtain a reduced-order LTI approximation that satisfies the optimality conditions. An outline of the method is given below.

For a random initialization of $\hat{A}_r,\hat{B}_r,\hat{C}_r$, the following pair of Sylvester equations 
 \begin{align*}
        \hat{A}X_{\tau}+X_{\tau}{\hat{A}_r}^{T}+\hat{B}{\hat{B}_r}^{T}-e^{\hat{A}\tau}\hat{B}{\hat{B}_r}^{T}e^{{\hat{A}_r}^{T}\tau} &= 0 \quad {and} \\
        \hat{A}^{T}Y_{\tau} + Y_{\tau}{\hat{A}_r}+\hat{C}^{T}{\hat{C}_r}-e^{\hat{A}^{T}\tau}\hat{C}^{T}\hat{C}_r e^{\hat{A}_r\tau} &= 0 
\end{align*}        
is solved. Then $\hat{V}_r, \hat{W}_r \in \mathbb{R}^{n \times r}$ are obtained such that $\text{Ran}(\hat{V}_r)= \text{Ran}(X_{\tau})$ and $\text{Ran}(\hat{W}_r)= \text{Ran}(Y_{\tau})$ with $\hat{W}_r$ and $\hat{V}_r$ having orthonormal columns. Further, $\hat{Z}_r^{T}=(\hat{W}_r^{T}\hat{V}_r)^{-1}\hat{W}_r^T$ is computed. An updated ROM is obtained as follows:
\begin{equation*}
        \hat{A}_r=\hat{Z}_r^{T}\hat{A}\hat{V}_r, \quad \hat{B}_r = \hat{Z}_r^{T}\hat{B}, \quad \text{and} \quad \hat{C}_r=\hat{C}\hat{V}_r.
\end{equation*}
These steps are repeated for the next iteration. The iterations are stopped when
the change in eigenvalues of $\hat{A}_r$ for two consecutive iterations becomes less than a preset tolerance. No formal proof of convergence is available. However, the algorithm has been tested using several examples and converges after a finite number of iterations.


\section{Finite horizon model order reduction for continuous-time LTV systems}\label{Section3}
The time-limited $H_2$ error norm and the TL-TSIA algorithm for LTI systems discussed in the previous section are extended to continuous-time LTV systems in this section.



Consider the following continuous-time LTV system,
\begin{align}
 \frac{dx}{dt}(t) &= A(t)x(t)+B(t)u(t),  \nonumber \\
  y(t) &= C(t)x(t).  \label{LTVsystem1}
\end{align}
 Here, $A(\cdot):[t_0,t_f] \to \mathbb{R}^{n \times n}$, $t \mapsto A(t)$, $B(\cdot):[t_0,t_f] \to \mathbb{R}^{n \times m}$, $t \mapsto B(t)$ and $C(\cdot):[t_0,t_f] \to \mathbb{R}^{p \times n}$, $t \mapsto C(t)$ are continuous and bounded. This ensures the existence and uniqueness of the solutions of the LTV system over $[t_0,t_f]$. Let $\phi(t,\tau)$ be the state transition matrix (STM), and $h(t,\tau)$ be the impulse response of the system. By definition, the state transition matrix satisfies the relation: $\phi(t,t)=I_n \forall t \in [t_0,t_f]$.
We now state a few properties of the STM \cite{kailath1980linear}. For all $t_i \in [t_0,t_f]$ where $i=1,2,3$,  
\begin{align}
     &a)\quad  \phi(t_1,t_2)=\phi(t_1,t_3)\phi(t_3,t_2),\quad  ,\label{STM_Prop2} \\
     &b)\quad \text{det}\left(\phi(t_1,t_2)\right) \neq 0, \quad \textrm{and} \label{STM_Prop3} \\ 
     &c)\quad \left(\phi(t_1,t_2)\right)^{-1}=\phi(t_2,t_1). \label{STM_Prop4}
\end{align}
The above STM properties are used for some of the proofs presented later in this section.

The state transition matrix $\phi(t,\tau)$ is also the unique solution of the differential equation
 \begin{equation}\label{DefnSTM}
     \frac{\partial}{\partial t}\phi(t,\tau) = A(t)\phi(t,\tau),
 \end{equation}
 with initial condition $\phi(\tau,\tau)=I_n$. The impulse response matrix of the LTV system is as follows: 
\begin{equation}\label{httau}
h(t,\tau)= \begin{cases}
          0, &  t_0 \leq t<\tau,\\
          C(t)\phi(t,\tau)B(\tau), &  \tau \leq t \leq t_f.
          \end{cases}
\end{equation}

\subsection{The finite horizon $H_2$ error norm}
We now introduce a finite horizon $H_2$ error norm for LTV systems. 
Minimizing this norm ensures that the output of the reduced-order LTV model is a good approximation of the original LTV model output over a finite time interval.

Consider the continuous-time LTV system $\Sigma_r$ of order $r$, where $r < n$:
\begin{align} 
\dot{x}_r(t) &= A_r(t) {x_r}(t)+B_r(t){u}(t), \nonumber \\
y_r(t) &= C_r(t) x_r(t).  \label{red_LTVsystem1}
\end{align}
Here, $A_r(\cdot):[t_0,t_f] \to \mathbb{R}^{n \times n}$, $t \mapsto A_r(t)$, $B_r(\cdot):[t_0,t_f] \to \mathbb{R}^{n \times m}$, $t \mapsto B_r(t)$ and $C_r(\cdot):[t_0,t_f] \to \mathbb{R}^{p \times n}$, $t \mapsto C_r(t)$ are continuous and bounded. If $\Sigma_r$ is a finite horizon reduced-order approximation of $\Sigma$, then it is expected that $y(t) \approx y_r(t)$ for $t \in [t_0,t_f]$. Let $\phi_r(t,\tau)$ and $h_r(t,\tau)$ be the state transition and impulse response matrix, respectively. 


For a permissible input $u(t)$, the output $y(t)$ of the full-order system $\Sigma$ is $y(t) = \int_{t_0}^{t}h(t,\tau)u(\tau)d\tau$.
For the same input $u(t)$, the output $y_r(t)$ of the reduced-order system $\Sigma_r$ is given by $y_r(t) = \int_{t_0}^{t}h_r(t,\tau)u(\tau)d\tau$. Taking the norm of the output error $e(t)=y(t)-y_r(t)$, we have
\begin{align}
  \left\Vert y(t)-y_r(t) \right\Vert_2 
      &= \left\Vert \int_{t_0}^{t} \left(h(t,\tau)-h_r(t,\tau)\right)u(\tau)d\tau \right\Vert_2 \nonumber \\ 
     &\leq \int_{t_0}^{t} \left\Vert h(t,\tau)-h_r(t,\tau) \right\Vert_F \left\Vert u(\tau) \right\Vert_2 d\tau. \nonumber 
\end{align}
Applying the Cauchy-Schwarz inequality to the right-hand side of the above expression gives
\begin{align}
&\left\Vert e(t) \right\Vert_2 
\leq \left(\int_{t_0}^{t}\left\Vert h(t,\tau)-h_r(t,\tau)\right\Vert_{F}^{2} d\tau\right)^{\frac{1}{2}}
\left\Vert u \right\Vert_{L_2^m[t_0,t_f]}. \nonumber   \\
&\left\Vert e \right\Vert_{L_2^p[t_0,t_f]} \leq \left(\int_{t_0}^{t_f}\int_{t_0}^{t}\left\Vert h(t,\tau)-h_r(t,\tau)\right\Vert_{F}^{2} d\tau dt \right)^{\frac{1}{2}} \left\Vert u \right\Vert_{L_2^m[t_0,t_f]}. \nonumber
\end{align}
Based on the above inequality, we observe that minimizing the term  $\left(\int_{t_0}^{t_f}\int_{t_0}^{t}\left\Vert h(t,\tau)-h_r(t,\tau) \right\Vert_{F} d\tau dt\right)^{\frac{1}{2}}$ ensures that the output error norm is minimized. This term is referred to as the finite horizon $H_2$ error norm and is denoted by $\left\Vert \Sigma-\Sigma_r \right\Vert_{H_2 [t_0,t_f]}$. 

The finite horizon $H_2$ optimal model order reduction problem for continuous-time LTV systems involves obtaining an approximation \eqref{red_LTVsystem1} of the original model \eqref{LTVsystem1} such that $\left\Vert \Sigma-\Sigma_r \right\Vert_{H_2 [t_0,t_f]}$ is minimized. We now define several terms used to obtain a trace formula for the error norm.  


Given initial time $t_0$, and a time-instant $t \in [t_0,t_f]$, the matrices $P(t)$, $P_r(t)$ and $X(t)$ are defined as
\begin{align}
&P(t) := \int_{t_0}^{t}\phi(t,\tau)B(\tau)(B(\tau))^{T}(\phi(t,\tau))^{T}d\tau, \label{IntExpP} \\
&P_r(t) := \int_{t_0}^{t}\phi_r(t,\tau)B_r(\tau)(B_r(\tau))^{T}(\phi_r(t,\tau))^{T}d\tau, \: \mathrm{and} \label{IntExpPr} \\ 
&X(t) := \int_{t_0}^{t}\phi(t,\tau)B(\tau)(B_r(\tau))^{T}(\phi_r(t,\tau))^{T}d\tau,  \label{IntExpX}
\end{align}
The above matrices 
can be computed by solving the following matrix differential equations from $t=t_0$ to $t=t_f$
\begin{align}
&\frac{dP(t)}{dt} = A(t)P(t)+P(t)(A(t))^{T}+B(t)(B(t))^{T}, \label{P_LTV}\\
&\frac{d}{dt}P_r(t) = A_r(t)P_r(t)+P_r(t)(A_r(t))^{T}+
B_r(t)(B_r(t))^{T}, \label{Pr_LTV}\\
&\frac{d}{dt}X(t) = A(t)X(t)+X(t)(A_r(t))^{T}+
B(t)(B_r(t))^{T},  \label{X_LTV} 
\end{align}
with $P(t_0)=0_{n \times n}$, $P_r(t_0)= 0_{r \times r}$ and $X(t_0)=0_{n \times r}$, respectively.

Similarly, given final time $t_f$ and a time-instant $t$, the matrices $Q(t)$, $Q_r(t)$ and $Y(t)$ are defined as follows
\begin{align}
&Q(t) := \int_{t}^{t_f}(\phi(\tau,t))^{T}(C(\tau))^{T}C(\tau)\phi(\tau,t)d\tau, \label{IntExpQ} \\ 
&Q_r(t) := \int_{t}^{t_f}(\phi_r(\tau,t))^{T}(C_r(\tau))^{T}C_r(\tau)\phi_r(\tau,t) d\tau,\label{IntExpQr} \\
&Y(t) := \int_{t}^{t_f} (\phi_r(\tau,t))^{T}(C_r(\tau))^{T}C(\tau)\phi(\tau,t)d\tau. \label{IntExpY}
\end{align}
The above matrices are computed by solving the following matrix differential equations from $t=t_f$ to $t=t_0$:
\begin{align}
&-\frac{dQ(t)}{dt} = (A(t))^{T}Q(t)+Q(t)A(t)+(C(t))^{T}C(t),
\label{Q_LTV} \\
&-\frac{d}{dt}Y(t) = (A(t))^{T}Y(t)+Y(t)A_r(t)+(C(t))^{T}C_r(t), \quad \text{and} \label{Y_LTV} \\
&-\frac{d}{dt}Q_r(t) = (A_r(t))^{T}Q_r(t)+Q_r(t)A_r(t)+(C_r(t))^{T}C_r(t), \label{Qr_LTV}
\end{align}
with $Q(t_f)=0_{n \times n}$, $Y(t_f)=0_{n \times r}$, and $Q_r(t_f)=0_{r \times r}$, respectively.

Among the matrices introduced here, $P(t)$ and $P_r(t)$
are the finite horizon reachability Gramians of $\Sigma$ \eqref{LTVsystem1} and $\Sigma_r$ \eqref{red_LTVsystem1}. Similarly, $Q(t)$ and $Q_r(t)$ are the finite horizon observability Gramians of $\Sigma$ and $\Sigma_r$, respectively. For a detailed discussion regarding the properties and uses of these Gramians, see \cite{kailath1980linear}.  
\begin{prop}
The square of the finite horizon $H_2$ error norm is expressed using  the reachability and the observability Gramians of $\Sigma$ and $\Sigma_r$ as follows,
\begin{align}
    &\left\Vert \Sigma-\Sigma_r \right\Vert_{H_2 [t_0,t_f]}^2  
    = \int_{t_0}^{t_f}\textrm{Tr}(C(t)P(t)(C(t))^{T}- \nonumber \\ 
    & 2C(t)X(t)(C_r(t))^{T}+ C_r(t)P_r(t)(C_r(t))^{T})dt.\label{ErrorCost1}\\
    &= \int_{t_0}^{t_f}\textrm{Tr}((B(t))^{T}Q(t)B(t)- 2(B(t))^{T}Y(t)B_r(t)+ \nonumber \\
    &(B_r(t))^{T}Q_r(t)B_r(t))dt. \label{ErrorCost2}
\end{align}
\end{prop}
\begin{pf}
The square of the error norm  $\left\Vert \Sigma-\Sigma_r \right\Vert_{H_2 [t_0,t_f]}^2$ involves doubles integration and can be expressed as follows:  
\begin{align}
   \left\Vert \Sigma-\Sigma_r \right\Vert_{H_2 [t_0,t_f]}^2 &= \int_{t_0}^{t_f} \int_{t_0}^{t} \left\Vert h(t,\tau)-h_r(t,\tau) \right\Vert_{F}^2 d\tau dt  \label{Cost Function1} \\
  &=\int_{t_0}^{t_f} \int_{\tau}^{t_f} \left\Vert h(t,\tau)-h_r(t,\tau) \right\Vert_{F}^2 dt d\tau. \label{Cost Function2} 
\end{align}
The integrand of the double integral in (\ref{Cost Function1}) is,
\begin{align}
    &\left\Vert h(t,\tau)-h_r(t,\tau) \right\Vert_{F}^2 \nonumber\\
    &=\mathrm{Tr}((C(t)\phi(t,\tau)B(\tau)-C_r(t)\phi_r(t,\tau)B_r(\tau))
    \times \nonumber \\
    &((B(\tau))^{T}(\phi(t,\tau))^{T}(C(t))^{T}-(B_r(\tau))^{T}({\phi}_r(t,\tau))^{T}(C_r(t))^{T})) \nonumber
\end{align}
Expanding the above expression and applying the double integral $\int_{t_0}^{t_f}\int_{t_0}^{t}(\cdot)d\tau dt$, we obtain the following terms
\begin{align}
    &\int_{t_0}^{t_f}\textrm{Tr}(C(t)( \int_{t_0}^{t}\phi(t,\tau)B(\tau)(B(\tau))^{T}(\phi(t,\tau))^{T}d\tau)\times \nonumber \\
    &(C(t))^{T})dt -2\int_{t_0}^{t_f}\mathrm{Tr}( C(t) (\int_{t_0}^{t}\phi(t,\tau)B(\tau)(B_{r}(\tau))^{T}\times \nonumber \\
    &(\phi_{r}(t,\tau))^{T}d\tau)(C_r(t))^{T})dt +
   \int_{t_0}^{t_f}\mathrm{Tr}( C_r(t) (\int_{t_0}^{t}\phi_r(t,\tau) \times \nonumber \\ 
   &B_r(\tau)(B_{r}(\tau))^{T}(\phi_{r}(t,\tau))^{T}d\tau) (C_r(t))^{T}) dt \nonumber
\end{align}
Substituting \eqref{IntExpP}, \eqref{IntExpX} and \eqref{IntExpPr} in the above expression, \eqref{ErrorCost1} is obtained. 

The integrand of the double integral in (\ref{Cost Function2}) becomes
\begin{align}
    &\left\Vert h(t,\tau)-h_r(t,\tau) \right\Vert_{F}^2 \nonumber\\
    &=\mathrm{Tr}(((B(\tau))^{T}(\phi(t,\tau))^{T}(C(t))^{T}-(B_r(\tau))^{T}({\phi}_r(t,\tau))^{T}\times \nonumber \\
    &(C_r(t))^{T})(C(t)\phi(t,\tau)B(\tau)-C_r(t)\phi_r(t,\tau)B_r(\tau))) \nonumber 
\end{align}
Expanding the above expression,  
applying the double integral $\int_{t_0}^{t_f}\int_{\tau}^{t_f}(\cdot)dt d\tau$ and following similar steps as the previous case results in \eqref{ErrorCost2}.
\end{pf}
\subsection{Functional derivatives of the finite horizon $H_2$ error norm}
The finite horizon $H_2$ optimal MOR problem for LTV systems discussed in the previous subsection is non-convex. Obtaining a global minimum for this problem is difficult. Hence, we derive conditions for optimality of the finite horizon $H_2$ error norm and try to attain reduced-order models satisfying these conditions.

First, we obtain the perturbation in the STM due to perturbation in the state matrix of an LTV system. 
\begin{lem}\label{PertSTMAt}
Consider the reduced-order LTV system given by \eqref{red_LTVsystem1}. Let the state matrix $A_r(t)$ be perturbed by a continuous mapping 
$\Delta A_r(t): [t_0,t_f] \to \mathbb{R}^{r \times r}$.
Let $(A_r(t)+\Delta A_r(t))$ be the perturbed state matrix of the reduced-order system and $\hat{\phi}_r(t,\tau)$ be the corresponding STM. The perturbation in the STM induced by the perturbation in the state matrix $\Delta \phi_r(t,t_0) = \hat{\phi}_r(t,t_0)-\phi_r(t,t_0)$ is as follows:
\begin{align}
    &\Delta \phi_r(t,t_0) = \int_{t_0}^{t}\phi_r(t,\tau)\Delta A_r(\tau)\phi_r(\tau,t_0)d\tau +  \nonumber \\
    &\int_{t_0}^{t}\int_{t_0}^{\tau}\phi_r(t,\tau)\Delta A_r(\tau)\phi_r(\tau,s)\Delta A_r(s)\hat{\phi}_r(s,t_0) ds \, d\tau. \label{trans_mat_per}
\end{align}
\end{lem}
\begin{pf}
For $u \equiv 0$, the solution of the LTV system $\Sigma_r$ given by (\ref{red_LTVsystem1}) becomes 
\begin{equation}\label{xrt}
    x_r(t) = \phi_r(t,t_0)x_r(t_0).
\end{equation}
Let $\hat{x}_r(t)$ be the state vector for the new state matrix $(A_r(t)+\Delta A_r(t))$. The new differential equation is as follows:
\begin{equation*}
\frac{d\hat{x}_r(t)}{dt} = (A_r(t)+\Delta A_r(t))\hat{x}_r(t).
\end{equation*}
Let $\hat{\phi}_r(t,s)$ be the STM for the new state matrix. For the same initial condition $x_r(t_0)$, the solution of the above differential equation is  
\begin{equation}\label{hatxrt}
    \hat{x}_r(t) = \hat{\phi}_r(t,t_0)x_r(t_0).
\end{equation}
Differentiating $\Delta x_r(t)=\hat{x}_r(t)-x_r(t)$ with respect to $t$ results in
\begin{align}
    \frac{d\Delta x_r(t)}{dt} &= \frac{d\hat{x}_r(t)}{dt}-\frac{dx_r(t)}{dt} \nonumber \\
    &= A_r(t)\Delta x_r(t) + \Delta A_r(t)\hat{x}_r(t). \nonumber 
\end{align}
For $t=t_0$, $\Delta x_r(t_0)=\hat{x}_r(t_0)-x_r(t_0)=0$. Thus, the solution of the above differential equation is,
\begin{align}
    \Delta x_r(t) &= \int_{t_0}^{t}\phi_r(t,\tau)\Delta A_r(\tau)\hat{x}_r(\tau) d\tau \nonumber \\
    &= \left(\int_{t_0}^{t}\phi_r(t,\tau)\Delta A_r(\tau)\hat{\phi}_r(\tau,t_0) d\tau \right) x_r(t_0). \label{Deltaxrt1}
\end{align}
Subtracting (\ref{hatxrt}) from (\ref{xrt}), gives the following expression for $\Delta x_r(t)$,
\begin{align}
    \Delta x_r(t) &= \left(\hat{\phi}_r(t,t_0)-\phi_r(t,t_0)\right)x_r(t_0) \nonumber \\
                  &= \Delta \phi_r(t,t_0) x_r(t_0). \label{Deltaxrt2}
\end{align}
Comparing (\ref{Deltaxrt1}) and (\ref{Deltaxrt2}) results in the following
\begin{align}
     \Delta \phi_r(t,t_0)x_r(t_0)=\left(\int_{t_0}^{t}\phi_r(t,\tau)\Delta A_r(\tau)\hat{\phi}_r(\tau,t_0) d\tau \right) x_r(t_0) . \nonumber 
\end{align}
Since the above equation holds for arbitrary $x_r(t_0)$, 
\begin{align}
&\Delta \phi_r(t,t_0) = \int_{t_0}^{t}\phi_r(t,\tau)\Delta A_r(\tau)\hat{\phi}_r(\tau,t_0) d\tau \label{Deltaphirtt01} \\
&= \int_{t_0}^{t}\phi_r(t,\tau)\Delta A_r(\tau){\phi}_r(\tau,t_0) d\tau  + \nonumber \\
&\int_{t_0}^{t}\phi_r(t,\tau)\Delta A_r(\tau) \Delta \phi_r(\tau,t_0) d\tau. \label{Deltaphirtt02}
\end{align}
From (\ref{Deltaphirtt01}), we get $\Delta \phi_r(\tau,t_0)=\int_{t_0}^{\tau}\phi_r(\tau,s)\Delta A_r(s)\hat{\phi}_r(s,t_0) ds$.  Using this expression in the second term on the right-hand side of (\ref{Deltaphirtt02}) gives
\begin{align}
    &\int_{t_0}^{t}\phi_r(t,\tau)\Delta A_r(\tau) \Delta \phi_r(\tau,t_0) d\tau  \nonumber \\
    &= \int_{t_0}^{t}\phi_r(t,\tau)\Delta A_r(\tau)\int_{t_0}^{\tau} \phi_r(\tau,s)\Delta A_r(s)\hat{\phi}_r(s,t_0) ds \, d\tau \nonumber \\
    &= \int_{t_0}^{t}\int_{t_0}^{\tau}\phi_r(t,\tau)\Delta A_r(\tau)\phi_r(\tau,s)\Delta A_r(s)\hat{\phi}_r(s,t_0) ds \, d\tau. \nonumber
\end{align}
Substituting the above expression in \eqref{Deltaphirtt02} gives \eqref{trans_mat_per}. This completes the proof of the lemma.
\end{pf}
For LTI systems, the perturbation expression obtained in the above lemma simplifies to the Frechet derivative term described in Proposition \ref{LyapbasedOptim}.
\begin{cor}
Let $\Delta_{1} \phi_r(t,t_0)=\int_{t_0}^{t}\phi_r(t,\tau)\Delta A_r(\tau)\phi_r(\tau,t_0)d\tau$. Let $L(e^{A_r (t-t_0)},\Delta A_r(t-t_0))$ be the Fr{\'e}chet derivative of the matrix exponential $e^{A_r t}$ along $\Delta A_r(t-t_0)$. If $A_r(t)=A_r$ and $\Delta A_r(t)=\Delta A_r$ $\forall t \in [t_0,t_f]$, then  $\Delta_1 \phi_r(t,t_0) = L(e^{A_r (t-t_0)},\Delta A_r(t-t_0))$.
\end{cor}
\begin{pf} 
For the given assumptions, we have
\begin{align}
&\Delta_1 \phi_r(t,t_0) 
=\int_{t_0}^{t}e^{A_r(t-\tau)}\Delta A_r e^{A_r(\tau-t_0)}d\tau \nonumber \\
&=\int_{0}^{t-t_0}e^{A_r(t-t_0-l)}\Delta A_r e^{A_r l}dl \nonumber \\
&= \int_{0}^{1}e^{A_r (t-t_0)(1-s)}\Delta A_r(t-t_0)e^{A_r(t-t_0)s}ds \nonumber \\
&=L(e^{A_r(t-t_0)},\Delta A_r(t-t_0)). \nonumber 
\end{align}
\end{pf}

Next, we consider the following definition.

Let $M_i=\{f_i|f_i:[t_0,t_f] \to \mathbb{R}^{m_i \times n_i} \mathrm{\:\:is \:\: continuous \:\: and \:\: bound
ed}\}$ for $i=1,2,\hdots,k$.
Consider $F$ as $F: M_1 \times M_2 \times \hdots \times M_k \to \mathbb{R}$. 
\begin{defn}[\cite{parr1989density}, Appendix A]
The functional derivative of $F$ with respect to $f_i \in M_i$ is a function given by $\frac{\partial F}{\partial f_i}:[t_0,t_f] \to \mathbb{R}^{m \times n}$ which satisfies
\begin{align}
    \left\langle \frac{\partial F}{\partial f_i},\Delta f_i \right\rangle &= \int_{t_0}^{t_f}\text{Tr}\left(\left(\frac{\partial F}{\partial f_i}(t)\right)^{T}\Delta f_i(t) \right)dt \nonumber \\
    &= \lim_{\epsilon \to 0}\frac{F[f_i+\epsilon \Delta f_i]-F[f_i]}{\epsilon},    
\end{align}
where $\epsilon$ is a scalar and $\Delta f_i:[t_0,t_f] \to \mathbb{R}^{m_i \times n_i}$ is an function in $M_i$.
\end{defn}

The above definition is used in Theorem \ref{FunctionalDerivativesLTV} to obtain functional derivatives of the finite horizon $H_2$ error norm. The derivatives are then used to obtain the conditions for optimality of the same error norm in Theorem \ref{LTVProj}.
\begin{thm}\label{FunctionalDerivativesLTV}
Consider $J[A_r, B_r, C_r]= \left\Vert \Sigma-\Sigma_r \right\Vert_{H_2[t_0,t_f]}^2$ where $A_r(t)$, $B_r(t)$ and $C_r(t)$ are given by \eqref{red_LTVsystem1}. The functional derivatives of $J$ with respect to $A_r(t)$, $B_r(t)$ and $C_r(t)$, respectively, are as follows:
\begin{align}
    \frac{\partial J}{\partial A_r}(t)  &=  
    2(Q_r(t)P_r(t)-(Y(t))^{T}X(t)), \label{GradJAt}\\
    \frac{\partial J}{\partial B_r}(t)  &= 2(Q_r(t)B_r(t)-(Y(t))^{T}B(t)) \label{GradJBt} \quad \textrm{and}\\
    \frac{\partial J}{\partial C_r}(t)  &= 
    2(C_r(t)P_r(t)-C(t)X(t)). \label{GradJCt}
\end{align}
\end{thm}
\begin{pf}
The inner product of $\frac{\partial J}{\partial B_r}$ and an arbitrary matrix-valued perturbation $\Delta B_r:[t_0,t_f] \to \mathbb{R}^{r \times m}$ is as follows:
\begin{align}
&\left\langle \frac{\partial J}{\partial B_r}, \Delta B_r \right\rangle = \int_{t_0}^{t_f}\mathrm{Tr}\left(\left(\frac{\partial J}{\partial B_r}(t)\right)^{T}\Delta B_r(t) \right)dt \label{InnerPdtBr} \\
&= \lim_{\epsilon \to 0+} \frac{1}{\epsilon}\left( J[A_r,B_r+\epsilon \Delta B_r ,C_r]-J[A_r,B_r,C_r] \right). \nonumber  
\end{align}
Substituting the expression of $J$ given by \eqref{ErrorCost2} in the above equation and using the identity $Tr(A^{T} B)=Tr(B^{T}A)$ results in
\begin{align}
&\lim_{\epsilon \to 0+} \frac{1}{\epsilon}\left( J[A_r,B_r+\epsilon \Delta B_r ,C_r]-J[A_r,B_r,C_r] \right) \nonumber \\
&= 2\int_{t_0}^{t_f} \mathrm{Tr}\left( \left(Q_r(\tau)B_r(\tau)-(Y(\tau))^{T}B(\tau) \right)^{T}\Delta B_r(\tau) \right)d\tau \nonumber \\
&= \langle 2\left(Q_r(\tau)B_r(\tau)-(Y(\tau))^{T}B(\tau) \right), \Delta B_r(\tau) \rangle. \nonumber  
\end{align}
Since $\Delta B_r$ is arbitrary, comparing the above expression with (\ref{InnerPdtBr}), \eqref{GradJBt} is obtained.

The inner product of $\frac{\partial J}{\partial C_r}$ and an arbitrary matrix-valued perturbation $\Delta C_r: [t_0,t_f] \to \mathbb{R}^{p \times r}$ is as follows
\begin{align}
&\left\langle \frac{\partial J}{\partial C_r}, \Delta C_r \right\rangle = \int_{t_0}^{t_f}\mathrm{Tr}\left(\left(\frac{\partial J}{\partial C_r}(t)\right)^{T}\Delta C_r(t) \right)dt \label{InnerPdtCr} \\
&= \lim_{\epsilon \to 0+} \frac{1}{\epsilon}\left( J[A_r,B_r,C_r+\epsilon \Delta C_r]-J[A_r,B_r,C_r] \right). \nonumber  
\end{align}
Considering $J$ given by (\ref{ErrorCost1}) in the above limit and using the identity $Tr(A^{T} B)=Tr(B^{T}A)$ gives  
\begin{align}
&\lim_{\epsilon \to 0+} \frac{1}{\epsilon}\left( J[A_r,B_r,C_r+\epsilon \Delta C_r]-J[A_r,B_r,C_r] \right) \nonumber \\
&= 2\int_{t_0}^{t_f}\mathrm{Tr}\left(\left(C_r(t)P_r(t) -C(t)X(t)\right)(\Delta C_r(t))^{T}  \right) dt \nonumber \\
&= 2\int_{t_0}^{t_f}\mathrm{Tr}\left(\left(C_r(t)P_r(t) -C(t)X(t)\right)^{T}\Delta C_r(t)  \right)dt. \nonumber 
\end{align}
Since $\Delta C_r$ is arbitrary, comparing the above expression with (\ref{InnerPdtCr}) results in \eqref{GradJCt}.

The inner product of $\frac{\partial J}{\partial A_r}$ and an arbitrary matrix-valued perturbation $\Delta A_r:[t_0,t_f] \to \mathbb{R}^{r \times r}$ is as follows
\begin{align}
&\left\langle \frac{\partial J}{\partial A_r}, \Delta A_r \right\rangle = \int_{t_0}^{t_f}\mathrm{Tr}\left(\left(\frac{\partial J}{\partial A_r}(t)\right)^{T}\Delta A_r(t) \right)dt \label{InnerPdtAr} \\
&= \lim_{\epsilon \to 0+} \frac{1}{\epsilon}\left( J[A_r+\epsilon \Delta A_r,B_r,C_r]-J[A_r,B_r,C_r] \right). \nonumber  
\end{align}
Let $A_r(t)$ be perturbed by $\epsilon \Delta A_r$. Therefore, by Equation (\ref{trans_mat_per}), we get the following
\begin{align}
    \Delta \phi_r(t,t_0) = 
     \epsilon\int_{t_0}^{t}\phi_r(t,\tau)\Delta A_r(\tau)\phi_r(\tau,t_0)d\tau+\epsilon^2\phi  \, . \label{trans_mat_per1}
\end{align}
where $\phi=\int_{t_0}^{t}\int_{t_0}^{\tau}\phi_r(t,\tau)\Delta A_r(\tau)\phi_r(\tau,s)\Delta A_r(s)\hat{\phi}_r(s,t_0) ds \, d\tau$. Considering the expression of $J$ given by Equation (\ref{ErrorCost1}), we get the following
\begin{align}
    &J[A_r+\epsilon \Delta A_r,B_r,C_r]-J[A_r,B_r,C_r] =\int_{t_0}^{t_f}\mathrm{Tr}(C_r(t)\times \nonumber \\ 
    &\Delta P_r(t)(C_r(t))^{T} )dt- 
    2\int_{t_0}^{t_f}\mathrm{Tr}\left(C(t)\Delta X(t) (C_r(t))^{T} \right)dt,  \label{PertJAt} 
\end{align}
where $\Delta P_r(t)$ and $\Delta X(t)$ are the perturbations in $P_r(t)$ and $X(t)$, respectively, due to the perturbation of the state matrix $A_r(t)$. Using the expression of $P_r(t)$ given by Equation (\ref{IntExpPr}), the first term in the right-hand side of the Equation (\ref{PertJAt}) can be simplified as follows,
\begin{align}
    &\int_{t_0}^{t_f} \mathrm{Tr}(C_r(t)\int_{t_0}^{t}\Delta\phi_r(t,\tau)B_r(\tau)(B_r(\tau))^{T}(\phi_r(t,\tau))^{T} d\tau (C_r(t))^{T}  \nonumber \\
    &+C_r(t)\int_{t_0}^{t}\phi_r(t,\tau)B_r(\tau)(B_r(\tau))^{T}(\Delta \phi_r(t,\tau))^{T} d\tau (C_r(t))^{T}+ \nonumber \\
    & C_r(t)\int_{t_0}^{t}\Delta \phi_r(t,\tau)B_r(\tau)(B_r(\tau))^{T}(\Delta \phi_r(t,\tau))^{T} d\tau (C_r(t))^{T} ) dt \nonumber \\
    &=2\int_{t_0}^{t_f} \mathrm{Tr}(C_r(t)\int_{t_0}^{t}\Delta\phi_r(t,\tau)B_r(\tau)(B_r(\tau))^{T}(\phi_r(t,\tau))^{T} d\tau \times \nonumber \\
    &(C_r(t))^{T}) dt+ \int_{t_0}^{t_f} \mathrm{Tr}(C_r(t)\int_{t_0}^{t}\Delta \phi_r(t,\tau)B_r(\tau)(B_r(\tau))^{T}\times \nonumber \\
    &(\Delta \phi_r(t,\tau))^{T} d\tau (C_r(t))^{T}) dt. \nonumber 
\end{align}
Substituting $\Delta \phi_r(t,\tau)$ from (\ref{trans_mat_per1}) in the above expression results in 
\begin{align}
&2\epsilon\int_{t_0}^{t_f}\mathrm{Tr}((C_r(t))^{T}C_r(t)\int_{t_0}^{t}\int_{\tau}^{t}\phi_r(t,s)\Delta A_r(s)\phi_r(s,\tau)B_r(\tau)\times \nonumber \\
&(B_r(\tau))^{T}(\phi_r(t,\tau))^{T} ds \, d\tau)dt + \mathrm{Tr}(\Psi), \label{trans_mat_per2} 
\end{align}
where $\mathrm{Tr}(\Psi)=\mathcal{O}(\epsilon^2)$. 

Consider the first term of the expression (\ref{trans_mat_per2}). Exchanging the order of integration of the variables $s$ and $\tau$ results in
\vspace{-5mm}
\begin{align}
    &2\epsilon\int_{t_0}^{t_f}\mathrm{Tr}((C_r(t))^{T}C_r(t)\int_{t_0}^{t}\int_{t_0}^{s}\phi_r(t,s)\Delta A_r(s)\phi_r(s,\tau)B_r(\tau) \times \nonumber \\
    &(B_r(\tau))^{T}(\phi_r(s,\tau))^{T}(\phi_r(t,s))^{T} d\tau \, ds  )dt \nonumber \\
    &= 2\epsilon\mathrm{Tr} \int_{t_0}^{t_f}\int_{t_0}^{t}(\phi_r(t,s))^{T}(C_r(t))^{T}C_r(t)\phi_r(t,s)\Delta A_r(s) \times\nonumber \\
    &\int_{t_0}^{s}\phi_r(s,\tau)B_r(\tau)(B_r(\tau))^{T}(\phi_r(s,\tau))^{T} d\tau \, ds \, dt \nonumber \\
    &= 2\epsilon\mathrm{Tr} \int_{t_0}^{t_f}\int_{t_0}^{t} (\phi_r(t,s))^{T}(C_r(t))^{T}C_r(t)\phi_r(t,s)\Delta A_r(s) \times \nonumber \\
    &P_r(s) ds \, dt. \nonumber 
\end{align}
Further, exchanging the order of integration of the variables $t$ and $s$ in the above expression yields
\begin{align}
   & 2\epsilon\mathrm{Tr} \int_{t_0}^{t_f} P_r(s) \left( \int_{s}^{t_f}(\phi_r(t,s))^{T}(C_r(t))^{T}C_r(t)\phi_r(t,s) dt\right)\times \nonumber \\
   &\Delta A_r(s) ds \nonumber \\
   &= 2\epsilon \int_{t_0}^{t_f}\mathrm{Tr} \left( \left(Q_r(t_f,s)P_r(s,t_0)\right)^{T}\Delta A_r(s) \right) ds. \nonumber 
\end{align}
The expression (\ref{trans_mat_per2}) can be simplified as
\begin{align}
    2\epsilon \int_{t_0}^{t_f}\mathrm{Tr} \left( \left(Q_r(t_f,s)P_r(s,t_0)\right)^{T}\Delta A_r(s) \right) ds+ \mathrm{Tr}(\Psi). \label{PertJAt1} 
\end{align}
Similarly, using $X(t)$ given by \eqref{IntExpX} and substituting $\Delta \phi_r(t,\tau)$ from (\ref{trans_mat_per1}), the second term in the right-hand side of Equation (\ref{PertJAt}) is simplified as follows
\begin{align}
    &  2\epsilon\mathrm{Tr} \int_{t_0}^{t_f} (C_r(t))^{T} C(t)\int_{t_0}^{t}\phi(t,\tau)B(\tau)(B_r(\tau))^{T}\int_{\tau}^{t}(\phi_r(s,\tau))^{T}\times \nonumber \\ 
    &(\Delta A_r(s))^{T}(\phi_r(t,s))^{T} ds \, d\tau \, dt +\mathrm{Tr}(\eta) \nonumber \\
     &= 2\epsilon\mathrm{Tr} \int_{t_0}^{t_f} (C_r(t))^{T} C(t)\int_{t_0}^{t}\int_{\tau}^{t}\phi(t,s)\phi(s,\tau)B(\tau)(B_r(\tau))^{T}\times \nonumber \\ 
     &(\phi_r(s,\tau))^{T} 
     (\Delta A_r(s))^{T}(\phi_r(t,s))^{T} ds \, d\tau \, dt +\mathrm{Tr}(\eta), \label{Trans_mat_per3} 
\end{align}
where $\mathrm{Tr}(\eta) = \mathcal{O}(\epsilon^2)$.
For the first term of the above expression, exchanging the order of integration of $s$ and $\tau$ results in
\begin{align}
&2\epsilon\mathrm{Tr} \int_{t_0}^{t_f} (C_r(t))^{T} C(t)\int_{t_0}^{t}\phi(t,s)\int_{t_0}^{s}\phi(s,\tau)B(\tau)(B_r(\tau))^{T}\times \nonumber \\
&(\phi_r(s,\tau))^{T} d\tau (\Delta A_r(s))^{T}(\phi_r(t,s))^{T} ds \, dt \nonumber \\
&=2\epsilon\mathrm{Tr} \int_{t_0}^{t_f}\int_{t_0}^{t} (\phi_r(t,s))^{T} (C_r(t))^{T} C(t)\phi(t,s)X(s)\times \nonumber \\
&(\Delta A_r(s))^{T} ds \, dt. \nonumber 
\end{align}
Further, changing the order of integration of $s$ and $t$ gives
\begin{align}
&2\epsilon\mathrm{Tr} \int_{t_0}^{t_f}\int_{s}^{t_f} (\phi_r(t,s))^{T} (C_r(t))^{T} C(t)\phi(t,s) dt X(s)(\Delta A_r(s))^{T} ds \nonumber \\
&= 2\epsilon\mathrm{Tr} \int_{t_0}^{t_f}\left(Y(s)X(s)\right)^{T}\Delta A_r(s) ds.  \nonumber 
\end{align}
Thus, the expression (\ref{Trans_mat_per3}) simplifies to 
\begin{align}
   2\epsilon\mathrm{Tr} \int_{t_0}^{t_f}\left(Y(s)X(s)\right)^{T}\Delta A_r(s) ds  + \mathrm{Tr}(\eta). \label{PertJAt2}
\end{align}
Substituting (\ref{PertJAt1}) and (\ref{PertJAt2}) in (\ref{PertJAt}) results in
\begin{align}
  &J[A_r+\epsilon \Delta A_r,B_r,C_r]-J[A_r,B_r,C_r] 
  = 2\epsilon\int_{t_0}^{t_f}\mathrm{Tr}((Q_r(s)P_r(s)-\nonumber \\
  &Y(s)X(s))^{T}\Delta A_r(s))ds+\mathrm{Tr}(\Psi-\eta), \nonumber   
\end{align}
where $\mathrm{Tr}(\Psi-\eta) = \mathcal{O}(\epsilon^2)$.
Dividing the above expression by $\epsilon$ and taking limit as $\epsilon \to 0$ yields
\begin{align}
&\lim_{\epsilon \to 0+} \frac{1}{\epsilon}\left(J[A_r+\epsilon \Delta A_r,B_r,C_r]-J[A_r,B_r,C_r]\right) \nonumber \\ 
&= 2\int_{t_0}^{t_f}\mathrm{Tr}(\left(Q_r(s)P_r(s)-Y(s)X(s)\right)^{T}\Delta A_r(s))ds. \nonumber
\end{align}
As $\Delta A_r(t)$ is arbitrary, comparing the above expression with (\ref{InnerPdtAr}) results in \eqref{GradJAt}.
\end{pf}



\begin{thm}\label{LTVProj}
Let the continuous time-varying matrices $A_r^*(t)$, $B_r^*(t)$ and $C_r^*(t)$ be a stationary point of the functional $J[A_r,B_r,C_r]$. Let $P_r^*(t)$, $Q_r^*(t)$, $X^*(t)$ and $Y^*(t)$ be the solutions of (\ref{Pr_LTV}), (\ref{Qr_LTV}), (\ref{X_LTV}) and (\ref{Y_LTV}), respectively, for $A_r(t)=A_r^*(t)$, $B_r(t)=B_r^*(t)$ and $C_r(t)=C_r^*(t)$. If $P_r^*(t)$ and $Q_r^*(t)$ are invertible at every instant $t \in [t_0,t_f]$, then 
\begin{align}
A_r^*(t) &= (W_r(t))^{T}\left(A(t)V_r(t)-\frac{dV_r(t)}{dt}\right), \quad \mathrm{or} \label{optim_Ar1} \\
             &= \left((W_r(t))^{T}A(t)+\frac{d}{dt}(W_r(t))^{T} \right)V_r(t),  \label{optim_Ar2} \\
B_r^*(t) &= (W_r(t))^{T}B(t),  \quad  \textrm{and} \label{optim_Br} \\
C_r^*(t) &= C(t)V_r(t), \label{optim_Cr}
\end{align}
where $(W_r(t))^{T}V_r(t)=I_r$ with $W_r(t)=Y^*(t)\left(Q_r^*(t)\right)^{-1}$ and $V_r(t)=X^*(t)\left(P_r^*(t)\right)^{-1}$. 
\end{thm}
\begin{pf}
Let ($A_r^*, B_r^*,C_r^*$) be a stationary point of the functional $J[A_r,B_r,C_r]$. Let us define $\frac{\partial J}{\partial A_r}^* := \left. \frac{ \partial J}{\partial A_r} \right\vert_{\left({A_r^*}, {B_r^*},{C_r^*}\right)}$, ${\frac{\partial J}{\partial B_r}}^{*}:=\left. \frac{ \partial J}{\partial B_r} \right\vert_{\left({A_r^*}, {B_r^*},{C_r^*}\right)}$ and ${\frac{\partial J}{\partial C_r}}^{*}:=\left. \frac{ \partial J}{\partial C_r} \right\vert_{\left({A_r^*}, {B_r^*},{C_r^*}\right)}$. For arbitrary $\Delta A_r$, $\Delta B_r$ and $\Delta C_r$ with appropriate dimensions, continuous over $[t_0,t_f]$, the following relations hold.
\begin{align}
    \left\langle {\frac{\partial J}{\partial A_r}}^*, \Delta A_r \right\rangle = \left\langle {\frac{\partial J}{\partial B_r}}^*, \Delta B_r \right\rangle = \left\langle {\frac{\partial J}{\partial C_r}}^*, \Delta C_r \right\rangle = 0. \nonumber\end{align}
As ${\frac{\partial J}{\partial A_r}}^*(t)$, ${\frac{\partial J}{\partial B_r}}^*(t)$ and  ${\frac{\partial J}{\partial C_r}}^*(t)$ are continuous in $[t_0,t_f]$, so for all $t \in [t_0,t_f]$ we have
\begin{align}
    &{\frac{\partial J}{\partial A_r}}^{*}(t) =  {Q_r^*}(t){P_r^*}(t)-(Y^*(t))^{T}{X^*}(t)  = 0, \label{OptimAt}\\
    &{\frac{\partial J}{\partial B_r}}^{*}(t) = {Q_r^*}(t){B_r^*}(t)-({Y^*}(t))^{T}B(t) = 0, \quad \text{and} \label{OptimBt}\\
   & {\frac{\partial J}{\partial C_r}}^{*}(t) = {C_r^*}(t){P_r^*}(t)-C(t){X^*}(t) = 0. \label{OptimCt}
\end{align}
From (\ref{OptimBt}) and (\ref{OptimCt}), we have 
\begin{align}
     {B_r^*}(t) 
               &= \left({Y^*}(t)\left({Q_r^*}(t)\right)^{-1}\right)^{T}B(t)
               = (W_r(t))^{T}B(t)\nonumber \quad \text{and} \\
    {C_r^*}(t) &= C(t){X^*}(t)\left({P_r^*}(t)\right)^{-1}=C(t)V_r(t), \nonumber 
\end{align}
where $W_r(t)={Y^*}(t)\left({Q_r^*}(t)\right)^{-1}$ and 
$V_r(t)={X^*}(t)\left({P_r^*}(t)\right)^{-1}$. Using (\ref{OptimAt}), we get
\begin{align}
     \left(W_r(t)\right)^{T}V_r(t) 
                                  &= I_r. \label{ProjExp} 
\end{align}
Left multiplying (\ref{X_LTV}) by $(W(t))^{T}$, substituting ${X^*}(t)=V_r(t){P_r^*}(t)$ and using (\ref{ProjExp}) gives
\begin{flalign}
    &(W_r(t))^{T}\frac{d}{dt}\left( V_r(t){P_r^*}(t) \right) = (W_r(t))^{T}A(t)V_r(t){P_r^*}(t)+ \nonumber \\
    &(W_r(t))^{T}V_r(t){P_r^*}(t)({A_r^*}(t))^{T}+ 
     (W_r(t))^{T}B(t)({B_r^*}(t))^{T}& \nonumber \\
    &\Rightarrow \frac{d}{dt} {P_r^*}(t) = \left( (W_r(t))^{T}A(t)V_r(t)- W_r(t))^{T}\frac{d}{dt} V_r(t)\right)
    {P_r^*}(t) \nonumber \\
    &+{P_r^*}(t)({A_r^*}(t))^{T}+{B_r^*}(t)({B_r^*}(t))^{T}.& \nonumber
\end{flalign}
Comparing the above matrix differential equation with (\ref{Pr_LTV}) results in
Equation \eqref{optim_Ar1}.
Similarly, taking transpose of (\ref{Y_LTV}), right multiplying it by $V(t)$, substituting ${Y^*}(t)=W_r(t){Q_r^*}(t)$, and using (\ref{ProjExp}) gives
\begin{align}
    -&\frac{d}{dt}({Q_r^*}(t))^{T} =({Q_r^*}(t))^{T}\left((W_r(t))^{T}A(t)+\frac{d}{dt}(W_r(t))^{T} \right)V(t) \nonumber \\
    &+({A_r^*}(t))^{T}({Q_r^*}(t))^{T}+({C_r^*}(t))^{T}{C_r^*}(t). \nonumber 
\end{align}
By comparing the above matrix differential equation with (\ref{Qr_LTV}), Equation \eqref{optim_Ar2} is obtained.
\end{pf}


\begin{rem}
 The error norm for the time-limited $H_2$ optimal MOR problem for LTI systems stated in Subsection \ref{Sec2Subsec2_1} differs from the finite horizon $H_2$ error norm for LTV systems. As the LTI problem assumes a reduced-order LTI approximation, the gradients of the error norm are time-invariant. Further, the finite time constraint of the problem and the time-invariant nature of the reduced-order model results in the matrix exponential and the Frechet derivative terms as mentioned in Subsection \ref{Sec2Subsec2_2}. The finite horizon MOR problem for LTV systems allows an LTV reduced-order model, leading to time-varying gradients, which we refer to as functional derivatives. Despite the finite time constraint, the reduced-order model's time-varying nature greatly simplifies the functional derivatives' expression. It is also the reason for the absence of the Frechet derivative term (as evident from Theorem \ref{FunctionalDerivativesLTV}).
\end{rem}

\subsection{A projection-based model reduction scheme}
Consider the equations \eqref{IntExpPr}, \eqref{IntExpX}, \eqref{IntExpQr}, \eqref{IntExpY} and \eqref{optim_Ar1},\eqref{optim_Br}, \eqref{optim_Cr}. They can be interpreted as two coupled equations, $(X,Y,P_r,Q_r)=f(A_r,B_r,C_r)$ and $(A_r,B_r,C_r)=g(X,Y,P_r,Q_r)$. Thus, if $(A_r,B_r,C_r)$ is a stationary point of $J[A_r,B_r,C_r]$, then it is also a fixed point as $g(f(A_r,B_r,C_r))=(A_r,B_r,C_r)$. This motivates the following iterative procedure: $(X, Y, P_r, Q_r)_{k+1}=f(A_r, B_r, C_r)_{k+1}$ and $(A_r, B_r, C_r)_{k+1}=g(X, Y, P_r, Q_r)_k$. If the starting (initial) point of the iterative procedure is near a fixed point, then it is supposed to converge to the fixed point.

Based on the above idea, a fixed-point iterative scheme for continuous-time LTV systems is proposed in Algorithm \ref{TSIA_LTVsystems}. An appropriate initial point is critical for the convergence of the iterative scheme. Hence, instead of random initialization, a reduced-order LTV model of order $r$, obtained by FH BT, is used to initialize the algorithm. 

Let $(A_r^1, B_r^1, C_r^1)$ be the initial reduced-order LTV model of order $r$. Let us define $\frac{\partial J}{\partial A_r}^1$ , ${\frac{\partial J}{\partial B_r}}^{1}$ and ${\frac{\partial J}{\partial C_r}}^{1}$ be the functional derivatives for $(A_r^1, B_r^1, C_r^1)$. Also, let $J_{A_r}^1:=\operatorname*{max}_{t \in [t_0,t_f]} \left\Vert \frac{\partial J}{\partial A_r}^1   \right\Vert_2$, $J_{B_r}^1:=\operatorname*{max}_{t \in [t_0,t_f]} \left\Vert \frac{\partial J}{\partial B_r}^1   \right\Vert_2$ and 
$J_{C_r}^1:=\operatorname*{max}_{t \in [t_0,t_f]} \left\Vert \frac{\partial J}{\partial C_r}^1   \right\Vert_2$. As the reduced-order model obtained by the FH BT algorithm is not optimal, $J_{A_r}^1$, $J_{B_r}^1$ and $J_{C_r}^1$ are non-zero. The following quantities are computed for every iteration of Algorithm \ref{TSIA_LTVsystems}.
\begin{align}
&\mathrm{DerAr}  := \left(\operatorname*{max}_{t \in [t_0,t_f]} \left\Vert  \frac{\partial J}{\partial A_r}(t)\right\Vert_2\right)  \textfractionsolidus  J_{A_r}^1,  \label{DerAr} \\
&\mathrm{DerBr} := \left(\operatorname*{max}_{t \in [t_0,t_f]}\left\Vert \frac{\partial J}{\partial B_r}(t) \right\Vert_2\right) \textfractionsolidus  J_{B_r}^1 \label{DerBr}, \mathrm{and} \\
&\mathrm{DerCr}  := \left(\operatorname*{max}_{t \in [t_0,t_f]}\left\Vert \frac{\partial J}{\partial C_r}(t) \right\Vert_2\right) \textfractionsolidus  J_{C_r}^1  \label{DerCr} 
\end{align}
In the above terms, normalization by $J_{A_r}^1$, $J_{B_r}^1$ and $J_{C_r}^1$ ensures the ease of comparing them for subsequent iterations.

\begin{algorithm}
   \caption{Finite horizon TSIA (FH TSIA) for LTV systems }
   \label{TSIA_LTVsystems}
   \KwIn{$A(t),B(t),C(t)$; A finite time-interval $[t_0,t_f]$; 
   Order of the reduced-order model ($r$); Initial $A_r^0(t),B_r^0(t),C_r^0(t)$, obtained by FH-BT over $[t_0,t_f]$;}
   \KwOut{$A_r(t),B_r(t),C_r(t)$;} 
 \While{($\textrm{not converged}^{*}$)}{
    1. Compute $X(t)$, $P_r(t)$, $Y(t)$ and $Q_r(t)$ by solving the matrix differential equations \eqref{X_LTV}, \eqref{Pr_LTV}, \eqref{Y_LTV} and \eqref{Qr_LTV} for updated values of $A_r(t)$, $B_r(t)$ and $C_r(t)$;\\
   2. Compute $\mathrm{DerAr}$, $\mathrm{DerBr}$, and $\mathrm{DerCr}$ as defined by \eqref{DerAr}, \eqref{DerBr} and \eqref{DerCr}, respectively;\\
     3. Compute the projection matrices: $V_r(t)=X(t)(P_r(t))^{-1}$ and 
    $W_r(t)=Y(t)\left(Q_r(t)\right)^{-1}$;\\ 
    4. Update the reduced-order matrices using Equations \eqref{optim_Ar1}, \eqref{optim_Br} and \eqref{optim_Cr};
}
$*$(The meaning of the term `not converged' is explained in Remark \ref{convergence}.)
\end{algorithm}

\begin{rem}\label{convergence}
Similar to the TL-TSIA algorithm for LTI systems, we do not have a convergence proof for the proposed FH TSIA algorithm. The algorithm is terminated when there is no significant change in the values of $\mathrm{DerAr}$, $\mathrm{DerBr}$ and $\mathrm{DerCr}$ for successive iterations. 
If $\mathrm{DerAr}$, $\mathrm{DerBr}$ and $\mathrm{DerCr}$ are less than $1$ at termination, it indicates the closeness of the obtained reduced-order model to the optimality conditions and an improvement over the FH BT algorithm. 
\end{rem}


We now discuss the computational complexity of the proposed FH TSIA algorithm and the existing FH BT algorithm. 
FH BT involves solving two differential equations with $n \times n$ matrix coefficients for the time interval $[t_0,t_f]$. Then, a pair of time-varying projection matrices are computed using Schur projection \cite{safonov1989schur,sandberg2006case}. In comparison, FH TSIA involves solving four differential equations, two with $n \times r$ and another two with $r \times r$ matrix coefficients for the same time interval. The solutions are used to construct a pair of time-varying projection matrices. The differential equations are solved using appropriate ordinary differential equation (ODE) solvers. Thus, the computational cost of FH BT and FH TSIA depends on the values of $n$ and $r$ and the time the ODE solver takes to solve the differential equations. Due to the greater number of matrix coefficients, the FH BT algorithm appears more computationally expensive. However, unlike FH BT, FH TSIA is an iterative algorithm. So, apart from the aforementioned factors, its computational cost also depends on the number of iterations required for convergence. Thus, FH BT is computationally expensive compared to FH TSIA, provided the latter requires fewer iterations for convergence.






\section{Numerical examples}\label{Section4}
 In this section, we demonstrate the convergence of FH TSIA with the help of two numerical examples. The accuracy of a reduced-order model (ROM) is determined by computing the output error norm ($\left\Vert y-y_r \right\Vert_{L_2[t_0,t_f]}$) for a unit step input. 
For ease of comparison, the output error norms of the ROMs obtained at various iterations of FH TSIA are scaled by the output error norm of the FH BT ROM  and denoted by $\left\Vert y_{e} \right\Vert_r$.

\subsection{\underline{Example 1:}} 
\begin{figure}[h]
  \begin{center}
    \includegraphics[width=0.4\textwidth]{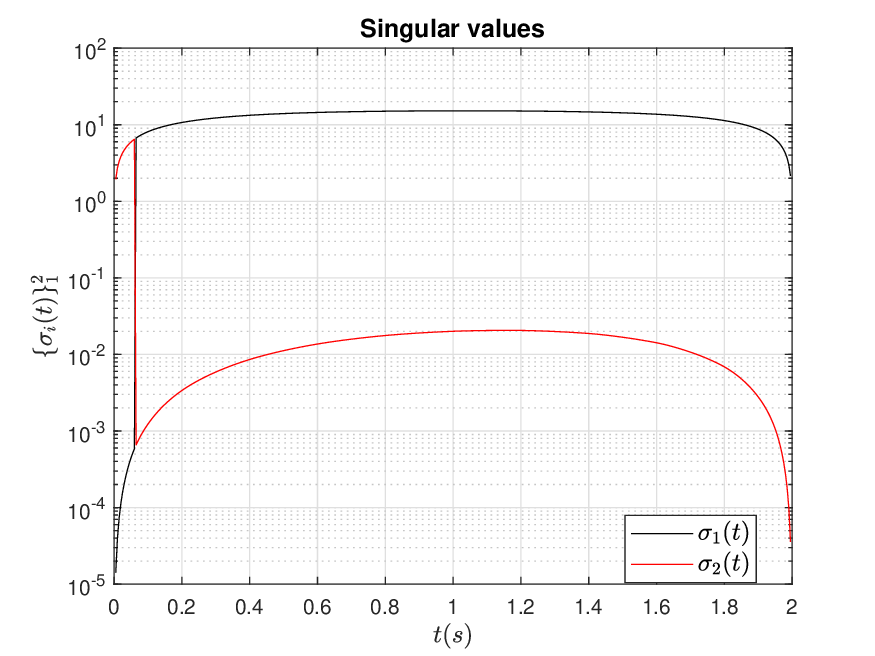}
    \caption{ Hankel singular values of the second-order LTV model.}
    \label{SingularValues_LTVsys}
  \end{center}
\end{figure}

The first example is an artificially constructed second-order LTV model. The parameters of the model are as follows:
\begin{align*}
A(t)=\begin{bmatrix}
t & 2e^{-t}\\1 & te^{-t}
\end{bmatrix},\: B(t)=\begin{bmatrix}
1 \\ 1 \end{bmatrix} \: \text{, and }  C(t) = \begin{bmatrix}
1 & 1 \end{bmatrix}.
\end{align*}
The finite horizon model order reduction problem is considered for the time interval $[0,2]$ $s$. Figure \ref{SingularValues_LTVsys} shows the singular values $\sigma_i(t)=\lambda_i^{\frac{1}{2}}(P(t)Q(t))$. The first singular value dominates the second singular value for almost the entire $[0,2]$ s, except the first $0.08$ s.

A first-order approximation of the LTV model is obtained using FH BT, which initialises FH TSIA. For this ROM, $J_{A_r}^1=7.56$, $J_{B_r}^1=136.95$, $J_{C_r}^1=147.09$ (defined in Subsection 3.3) and the output error norm is $0.35$. From Table \ref{TableLTVsys}, we see that there is no significant change in values of $\mathrm{DerAr}$, $\mathrm{DerBr}$, $\mathrm{DerCr}$ and $\left\Vert y_e\right\Vert_{r}$ beyond the sixth iteration. So, FH TSIA is terminated at the sixth iteration. The derivative and output error values indicate that the FH TSIA ROM is closer to optimality than the FH BT ROM.
\begin{table}[h]
\begin{center}
\begin{tabular}{|p{1.2cm}|p{0.5cm}|p{0.8cm}|p{0.8cm}|p{0.8cm}|p{0.8cm}|} \hline
Reduced-order &Iter  & ${\mathrm{DerAr}}$ & ${\mathrm{DerBr}}$ & ${\mathrm{DerCr}}$ & $\left\Vert y_e\right\Vert_{r}$ \\ \hline
\multirow{6}{*}{$r=1$}  & $1$  & $1.54$ & $1.02$ & $1.03$ & $1.33$ \\ 
  & $2$  & $1.48$ & $1.02$ & $1.04$ & $1.31$ \\ 
  & $3$  & $1.48 $ & $0.99$ & $1.03$ & $2$ \\ 
  & $4$  & $3.37$ & $1.06$ & $1.09$ & $1.38$ \\ 
  & $5$  & $0.63$ & $1.04$ & $1.09 $ & $0.41$ \\ 
  & $6$  & $0.42$ & $1.04$ & $1.10$ & $0.40$ \\ \hline
\end{tabular}
\caption{$\mathrm{DerAr}$, $\mathrm{DerBr}$, $\mathrm{DerCr}$ and $\left\Vert y_e\right\Vert_{r}$ for the first-order ROMs corresponding to iterations of FH TSIA.}
\label{TableLTVsys}
\end{center}
\end{table}

\begin{figure}[h]
  \begin{center}
    \includegraphics[width=0.4\textwidth]{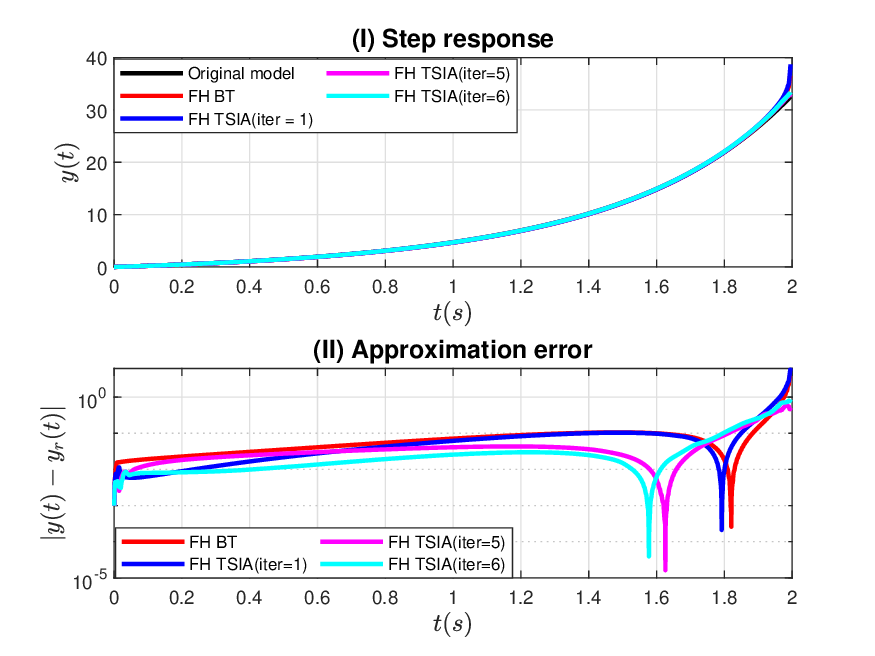}
    \caption{ (I) Step responses for the second-order LTV model and the first-order LTV approximations achieved by FH BT and various iterations of FH TSIA, (II) Approximation errors between the step responses of the original and the reduced-order models.}
    \label{Plots_LTVsys}
  \end{center}
\end{figure}

Figure \ref{Plots_LTVsys} compares the step responses of the original LTV model and the reduced-order approximations and the absolute output errors between them. It also indicates that among the FH TSIA and FH BT ROMs, the unit step response of the former is closer to the unit step response of the original model.

\subsection{\underline{Example 2:}} 
\begin{figure}[h]
  \centering
    \includegraphics[width=0.4\textwidth]{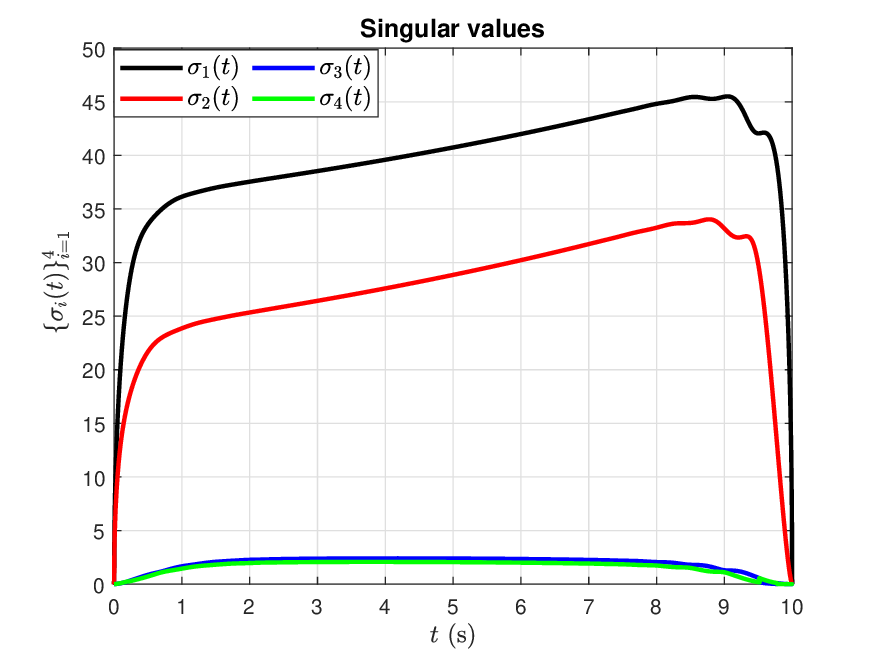}
    \caption{ Hankel singular values of the fourth-order LTV model.}
    \label{SingularValues_BTT}
\end{figure}

The second example is a fourth-order LTV model of a missile's pitch/yaw channel. The inputs are the actuator deflections $\delta_y (t)$ and $\delta_z (t)$; the output is the yaw rate $\omega_y (t)$. A detailed description of the model and its parameters is available in \cite{tan2016finite}. The finite horizon MOR problem is considered for the time interval $[0,10]$. Figure \ref{SingularValues_BTT} shows that the first ($\sigma_1(t)$) and the second ($\sigma_2(t)$) singular values dominate the third ($\sigma_3(t)$) and fourth ($\sigma_4(t)$) singular values.
\begin{figure}[h]
  \begin{center}
    \includegraphics[width=0.4\textwidth]{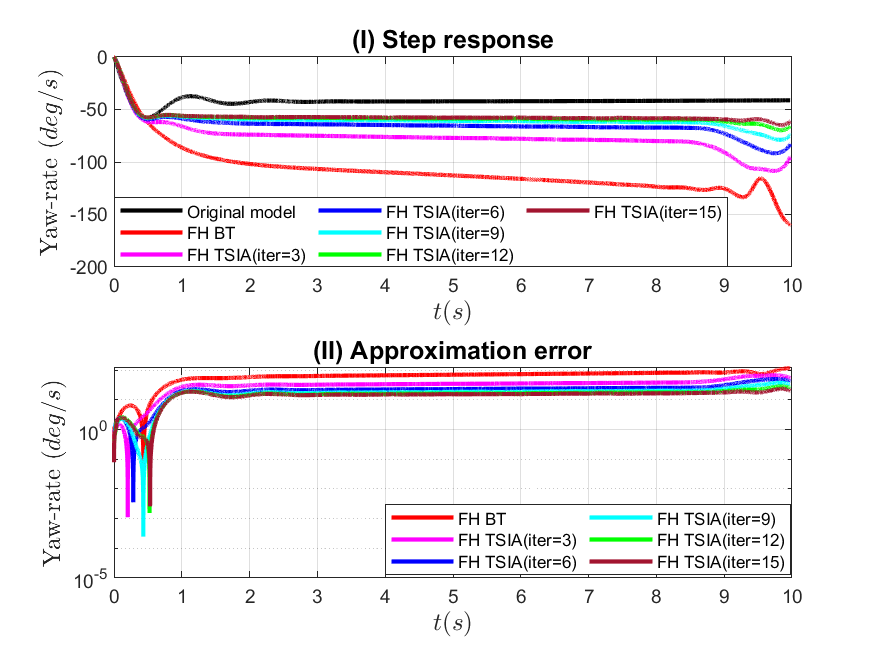}
    \caption{(I) Step responses for the $4^{th}$-order LTV model and the first-order approximations obtained by FH BT and various iterations of FH TSIA. (II) Approximation errors between the step responses of the original and the reduced-order models.}
    \label{BTT_Ord1}
   \end{center} 
\end{figure}

\begin{figure}[h]
  \begin{center}
    \includegraphics[width=0.4\textwidth]{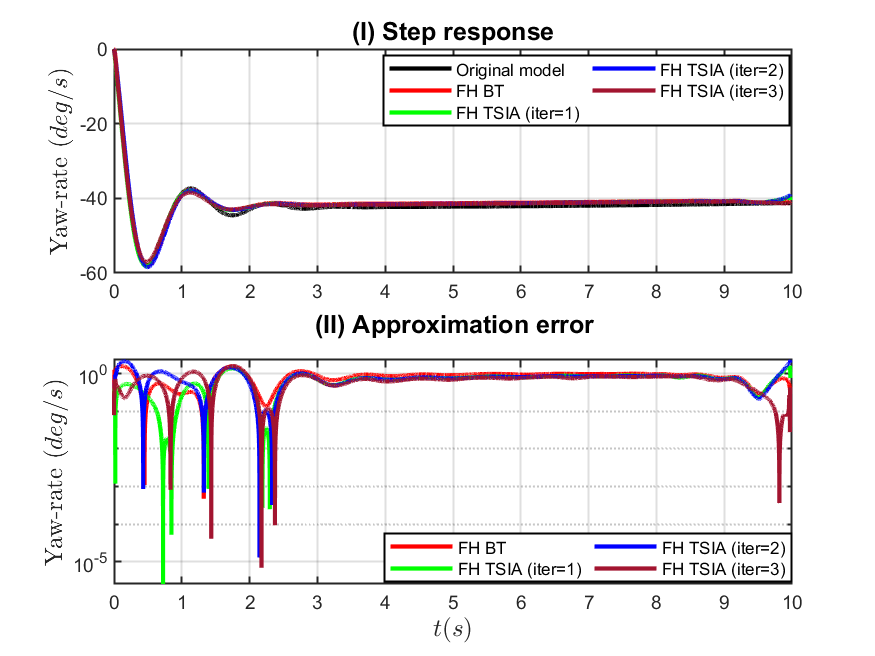}
    \caption{ (I) Step responses for the $4^{th}$-order LTV model and the second-order LTV approximations obtained by FH BT and various iterations of FH TSIA. (II) Absolute value of error between the step responses of the original and the reduced-order models. }
    \label{BTT_Ord2}
   \end{center} 
\end{figure}

A first-order approximation of the LTV model is obtained using FH BT. For the first-order ROM, $J_{A_r}^1=3937.9$, $J_{B_r}^1=918.7$, $J_{C_r}^1=748.8$ and the output error norm is $220.9$. This model is used to initialize FH TSIA. Table \ref{TableBTTord1} shows that $\mathrm{DerAr}$, $\mathrm{DerBr}$ and $\mathrm{DerCr}$ and $\left\Vert y_e \right\Vert_r$ do not change significantly beyond the $15^{th}$ iteration. Hence, FH TSIA is terminated at the $15^{th}$ iteration. Step responses are applied at both inputs of the original and the reduced-order models. The outputs are compared in Figure \ref{BTT_Ord1}. The FH TSIA ROM is much closer to optimality and has a considerably lesser output error norm than the FH BT ROM.
\begin{table}[h]
\begin{center}
\begin{tabular}{|p{1.2cm}|p{0.5cm}|p{0.8cm}|p{0.8cm}|p{0.8cm}|p{0.8cm}|} \hline
Reduced-order &Iter  & ${\mathrm{DerAr}}$ & ${\mathrm{DerBr}}$ & ${\mathrm{DerCr}}$ & $\left\Vert y_e\right\Vert_{r}$ \\ \hline
\multirow{5}{*}{$r=1$}  & $3$  & $0.73$ & $0.76$ & $1.43$ & $0.53$ \\ 
  & $6$  & $0.43$ & $0.76$ & $1.60$ & $0.35$ \\ 
  & $9$  & $0.22$ & $0.76$ & $1.68$ & $0.28$ \\ 
  & $12$  & $0.09$ & $0.75$ & $1.60$ & $0.24$ \\ 
  & $15$  & $0.03$ & $0.75$ & $1.60$ & $0.22$ \\ \hline
\multirow{3}{*}{$r=2$}  & $1$  & $2.51$ & $0.32$ & $1$ & $0.84$ \\ 
  & $2$  & $2.10$ & $0.13$ & $1$ & $0.96$ \\ 
  & $3$  & $1.05$ & $0.11$ & $0.79$ & $0.83$ \\ \hline
\end{tabular}
\caption{$\mathrm{DerAr}$, $\mathrm{DerBr}$, $\mathrm{DerCr}$ and $\left\Vert y_e\right\Vert_{r}$ for various iterations of FH TSIA.}
\label{TableBTTord1}
\end{center}
\end{table}

Then, a second-order approximation of the LTV model is obtained using FH BT.
For this model, $J_{A_r}^1=29.0$, $J_{B_r}^1=317.2$, $J_{C_r}^1=3.21 \times 10^{-13}$ and the output error norm is $2.8$. This shows that second-order FH BT ROM is a much better approximation of the original model than the first-order ROM. This model is then used to initialize FH TSIA, which converges in just three iterations, as seen from Table \ref{TableBTTord1}. Compared to the significant improvement of the FH TSIA ROM for the first-order approximation, there is only a slight improvement for the second-order approximation. This is because the second-order FH BT ROM is already close to optimality. Figure \ref{BTT_Ord2} displays the step responses of the original and the second-order ROMs and the absolute errors between them.

For both numerical examples, we observe that FH TSIA converges after a finite number of iterations and performs better than FH BT.

\section{Conclusion}\label{Section5}
This paper has proposed a finite horizon $H_2$ error norm for continuous-time LTV models. The functional derivatives of the proposed error norm have been derived. Further, they have been used to obtain conditions for optimality of the finite horizon $H_2$ error norm. Finally, a projection-based iterative scheme for model reduction of continuous-time LTV models has been proposed based on the optimality conditions. The performance of the iterative scheme has been illustrated via two numerical examples.

\bibliographystyle{plain}        
\bibliography{autosam}           

\begin{thebibliography}{10}

\bibitem{chahlaoui2005model}
Y.~Chahlaoui and P.~Van~Dooren.
\newblock Model reduction of time-varying systems.
\newblock In {\em Dimension Reduction of Large-Scale Systems: Proceedings of a
  Workshop held in Oberwolfach, Germany, October 19--25, 2003}, pages 131--148.
  Springer, 2005.

\bibitem{das2022h}
K.~Das, S.~Krishnaswamy, and S.~Majhi.
\newblock {$H_2$} optimal model order reduction over a finite time interval.
\newblock {\em IEEE Control Systems Letters}, 6:2467--2472, 2022.

\bibitem{das2023near}
K.~Das, S.~Krishnaswamy, and S.~Majhi.
\newblock Near-optimal interpolation-based time-limited model order reduction.
\newblock {\em International Journal of Control}, pages 1--11, 2023.

\bibitem{evers1992ltv}
J.~Evers, J.~Cloutier, and C.~Lin.
\newblock A {LTV} dynamics model for missile guidance and control in the
  endgame.
\newblock In {\em Astrodynamics Conference}, page 4533, 1992.

\bibitem{gawronski1990model}
W.~Gawronski and J.~Juang.
\newblock Model reduction in limited time and frequency intervals.
\newblock {\em International Journal of Systems Science}, 21(2):349--376, 1990.

\bibitem{goyal2019time}
P.~Goyal and M.~Redmann.
\newblock Time-limited ${H}_2$-optimal model order reduction.
\newblock {\em Applied Mathematics and Computation}, 355:184--197, 2019.

\bibitem{kailath1980linear}
T.~Kailath.
\newblock {\em Linear systems}, volume 156.
\newblock Prentice-Hall Englewood Cliffs, NJ, 1980.

\bibitem{lall2003error}
S.~Lall and C.~Beck.
\newblock Error-bounds for balanced model-reduction of linear time-varying
  systems.
\newblock {\em IEEE Transactions on Automatic Control}, 48(6):946--956, 2003.

\bibitem{lang2016balanced}
N.~Lang, J.~Saak, and T.~Stykel.
\newblock Balanced truncation model reduction for linear time-varying systems.
\newblock {\em Mathematical and Computer Modelling of Dynamical Systems},
  22(4):267--281, 2016.

\bibitem{melchior2012finite}
S.~A. Melchior, P.~Van~Dooren, and K.~A. Gallivan.
\newblock Finite horizon approximation of linear time-varying systems.
\newblock {\em IFAC Proceedings Volumes}, 45(16):734--738, 2012.

\bibitem{melchior2014model}
S.~A. Melchior, P.~Van~Dooren, and K.~A. Gallivan.
\newblock Model reduction of linear time-varying systems over finite horizons.
\newblock {\em Applied Numerical Mathematics}, 77:72--81, 2014.

\bibitem{parr1989density}
R.~G. Parr and W.~Yang.
\newblock {\em Density Functional Theory of Atoms and Molecules}.
\newblock New York: Oxford Univ Press, 1989.

\bibitem{safonov1989schur}
M.~G. Safonov and R.~Y. Chiang.
\newblock A {S}chur method for balanced-truncation model reduction.
\newblock {\em IEEE Transactions on automatic control}, 34(7):729--733, 1989.

\bibitem{sandberg2006case}
H.~Sandberg.
\newblock A case study in model reduction of linear time-varying systems.
\newblock {\em Automatica}, 42(3):467--472, 2006.

\bibitem{sandberg2004balanced}
H.~Sandberg and A.~Rantzer.
\newblock Balanced truncation of linear time-varying systems.
\newblock {\em IEEE Transactions on Automatic Control}, 49(2):217--229, 2004.

\bibitem{sinani2019h2}
K.~Sinani and S.~Gugercin.
\newblock $\mathcal{H}_{2}(t_{f})$ optimality conditions for a finite-time
  horizon.
\newblock {\em Automatica}, 110:108604, 2019.

\bibitem{tan2016finite}
F.~Tan, B.~Zhou, and GR~Duan.
\newblock Finite-time stabilization of linear time-varying systems by piecewise
  constant feedback.
\newblock {\em Automatica}, 68:277--285, 2016.

\bibitem{verriest1983generalized}
E.~Verriest and T.~Kailath.
\newblock On generalized balanced realizations.
\newblock {\em IEEE Transactions on Automatic Control}, 28(8):833--844, 1983.

\end{thebibliography}

\end{document}